# On the accessibility of adaptive phenotypes of a bacterial metabolic network

Wilfred Ndifon[1,*], Joshua B. Plotkin[2], Jonathan Dushoff[3]

[1]Department of Ecology and Evolutionary Biology, Princeton University, Princeton, NJ

[2]Department of Biology and Program in Applied Mathematics and Computational Science, University of Pennsylvania, Philadelphia, PA

[3]Department of Biology, McMaster University, Hamilton, ON, Canada

## Abstract

The mechanisms by which adaptive phenotypes spread within an evolving population after their emergence are understood fairly well. Much less is known about the factors that influence the evolutionary accessibility of such phenotypes, a pre-requisite for their emergence in a population. Here, we investigate the influence of environmental quality on the accessibility of adaptive phenotypes of *Escherichia coli*'s central metabolic network. We used an established flux-balance model of metabolism as the basis for a genotype-phenotype map (GPM). We quantified the effects of seven qualitatively different environments (corresponding to both carbohydrate and gluconeogenic metabolic substrates) on the structure of this GPM. We found that the GPM has a more rugged structure in qualitatively poorer environments, suggesting that adaptive phenotypes could be intrinsically less accessible in such environments. Nevertheless, on average ~74% of the genotype can be altered by neutral drift, in the environment where the GPM is most rugged; this could allow evolving populations to circumvent such ruggedness.

[*] Email address for correspondence: ndifon@gmail.com

Furthermore, we found that the normalized mutual information (NMI) of genotype differences relative to phenotype differences, which measures the GPM's capacity to transmit information about phenotype differences, is positively correlated with (simulation-based) estimates of the accessibility of adaptive phenotypes in different environments. These results are consistent with the predictions of a simple analytic theory that makes explicit the relationship between the NMI and the speed of adaptation. The results suggest an intuitive information-theoretic principle for evolutionary adaptation; adaptation could be faster in environments where the GPM has a greater capacity to transmit information about phenotype differences. More generally, our results provide insight into fundamental environment-specific differences in the accessibility of adaptive phenotypes, and they suggest opportunities for research at the interface between information theory and evolutionary biology.

Keywords: Adaptation; genotype-phenotype map; fitness landscape; evolvability; information theory

## Author Summary

Adaptation involves the discovery by mutation and spread through populations of traits (or "phenotypes") that have high fitness under prevailing environmental conditions. While the spread of adaptive phenotypes through populations is mediated by natural selection, the likelihood of their discovery by mutation depends primarily on the relationship between genetic information and phenotypes (the genotype-phenotype mapping, or GPM). Elucidating the factors that influence the structure of the GPM is therefore critical to understanding the adaptation process. We investigated the influence of environmental quality on GPM structure for a well-studied model of *Escherichia coli*'s metabolism. Our results suggest that the GPM is more rugged in qualitatively poorer environments and, therefore, the discovery of adaptive phenotypes may be intrinsically less likely in such environments. Nevertheless, we found that the GPM contains large neutral networks in all studied environments, suggesting that populations adapting to these environments could circumvent the frequent "hill descents" that would otherwise be required by a rugged GPM. Moreover, we demonstrated that adaptation proceeds faster in environments for which the GPM transmits information about phenotype differences more efficiently, providing a connection between information theory and evolutionary theory. These results have implications for understanding constraints on adaptation in nature.

## Introduction

During adaptation, a population "moves" in genotype space in search of genotypes associated with high-fitness phenotypes. The success of adaptation depends crucially on the accessibility of such adaptive phenotypes. While adaptive phenotypes rely on natural selection for their fixation, their accessibility depends, primarily, on the structure of the genotype to phenotype mapping (GPM) and, secondarily, on the forces that move a population in genotype space – i.e. selection and genetic drift (see Materials and Methods for relevant definitions). In particular, accessible phenotypes must be linked by a path of viable phenotypes to the initial phenotype of a population. In addition, the structure of the GPM determines the dominant mechanism by which a population moves in genotype space; for a smooth GPM that contains extensive neutral networks of genotypes associated with individual adaptive phenotypes, the motion may occur predominantly by genetic drift, with selection acting only occasionally to move a population from one neutral network onto another [1-3]. On the other hand, for very rugged GPMs having a low degree of neutrality, movement in genotype space may be mostly mediated by selection.

By studying the factors that influence biologically relevant GPMs, we may gain insight into the accessibility of adaptive phenotypes. To that end, we have taken advantage of recent advances in the understanding of bacterial metabolic networks [4-8] to investigate the influence of environmental quality on the structure of *E. coli*'s central metabolic network GPM [9,10] (see Table S1). We used the latest gene-protein reaction-associations data on the metabolic network [10] to identify all the genes involved in the network's central metabolic pathways (i.e., respiration, the tricarboxylic acid cycle, glycogen/gluconeogenesis, pyruvate metabolism, the pentose shunt). We found 166 such genes (see Table S1), and defined the network's genome to be an ordered list of these genes. Mutations to a given gene are allowed to change the gene's state from "on" to "off" (deleterious mutations) and vice-versa (compensatory mutations). A genotype is defined as a particular configuration of on-off states of the 166 genes that make up the genome. The Hamming distance between any two genotypes is the number of differences in the states of corresponding genes.

We define a genotype's phenotype (equivalent, for our purposes, to fitness) using a model of metabolic flux. Specifically, a growing body of experimental and theoretical work [7,11-15] suggests that under conditions of carbon limitation, *E. coli* (and other bacteria) organize their metabolic fluxes so as to optimize the production of biomass, and that experimentally realized optimal biomass yields can be predicted with reasonable accuracy by the mathematical method of flux-balance analysis [4-7]. Therefore, we used the optimal biomass yield predicted by means of flux-balance analysis as a biologically grounded proxy for the phenotype/fitness of a particular genotype of the metabolic network (Further details on the definition of the metabolic network GPM are given in Materials and Methods). We then used statistical and information-theoretic methods to investigate the structure of the GPM under conditions in which one of seven compounds (henceforth called “environments”) served as the primary metabolic substrate. Note that an advantage to studying *E. coli*’s central metabolic network is that its GPM is systemic (as are organismal GPMs), and it has a very rich structure; the network phenotype is an emergent property of interactions among gene products and between these products and the intracellular and extracellular environments of the bacterium. The interaction rules are numerous (i.e., of the order of the number of genes) and, in some cases, complex (see Figure 1).

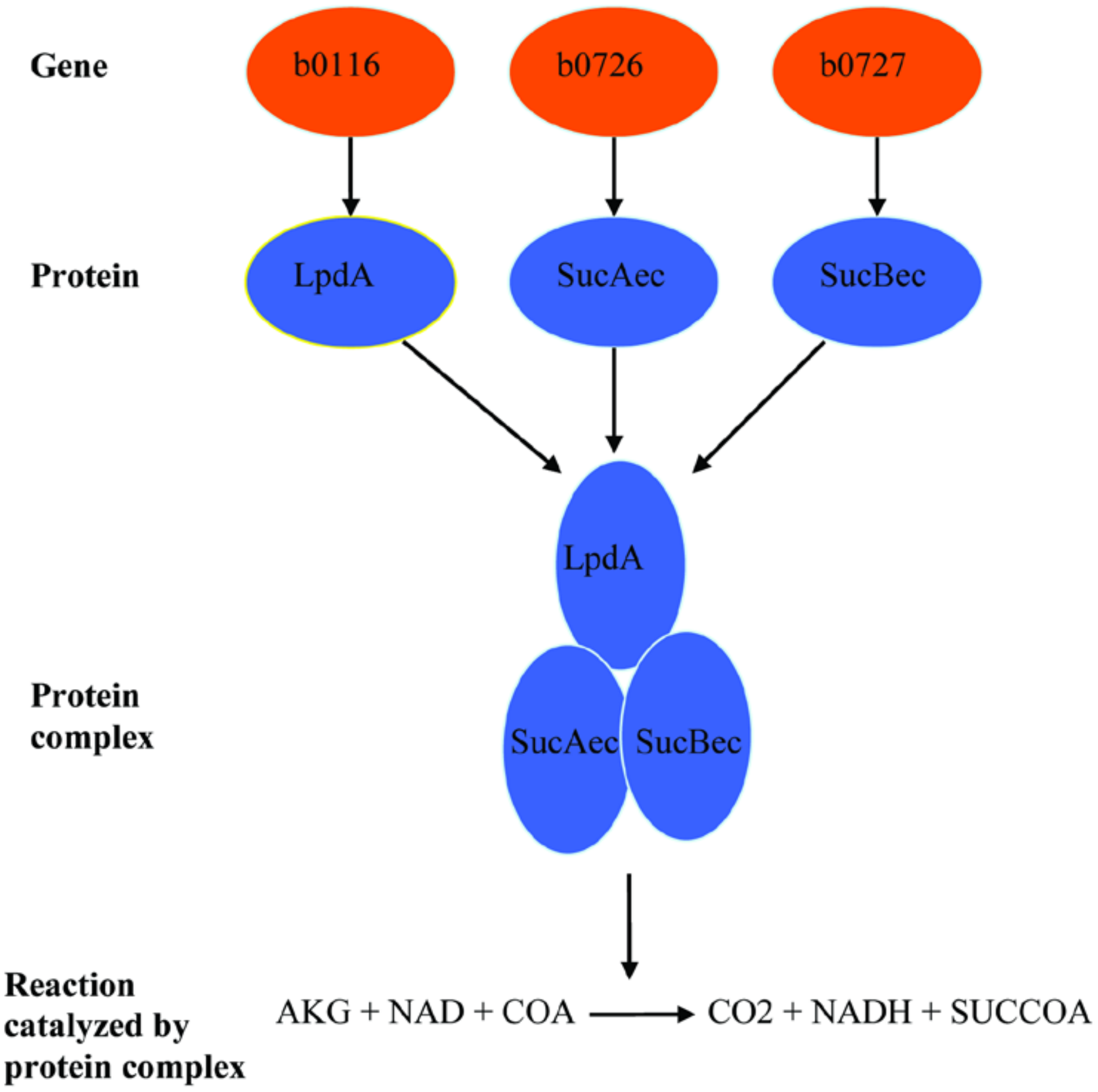

$$S = \begin{bmatrix} -1 & 1 & 0 & 0 & 0 & 0 & 0 \\ -1 & 0 & 1 & 0 & 0 & 0 & 0 \\ -1 & 0 & 0 & 1 & 0 & 0 & 0 \\ 1 & 0 & 0 & 0 & -1 & 0 & 0 \\ 1 & 0 & 0 & 0 & 0 & -1 & 0 \\ 1 & 0 & 0 & 0 & 0 & 0 & -1 \end{bmatrix}, \quad V = \begin{bmatrix} v \\ b_1 \\ b_2 \\ b_3 \\ b_4 \\ b_5 \\ b_6 \end{bmatrix}$$

Figure 1. **An example of the interaction rules found in the *E. coli* metabolic network**. The protein products of the genes b0116, b0726, and b0727 combine to form a protein complex that catalyzes production of succinate coenzyme A (SUCCOA) from alpha-ketoglutarate (AKG) and coenzyme A, with the concomitant reduction of nicotinamide adenine dinucleotide (NAD) and release of carbon dioxide ($CO_2$). A matrix $S$ of the stoichiometries of the reactants, and a vector $V$

of fluxes are shown. $v$, $b_1$, $b_2$, $b_3$, $b_4$, $b_5$, and $b_6$ denote the rates of the above reaction, the production of AKG, NAD, and COA, and the utilization of $CO_2$, NADH, and SUCCOA, respectively (Note that this is a simplification of the way the reaction is actually represented in our model). At steady state $S \cdot V=0$. In the event that one of the genes catalyzing the above reaction is turned off by mutation, the reaction flux $v$ is set to 0. Abbreviations (gene/protein product): b0116/LpdA, dihydrolipoamide dehydrogenase; b0726/SucAec, alpha-ketoglutarate decarboxylase; b0727/SucBec, dihydrolipoamide acetyltransferase.

## Results

Below, we describe the results of our analyses of the influence of the environment on aspects of the structure of the *E. coli* central metabolic network GPM that are important for the evolutionary accessibility of adaptive phenotypes. We used a flux-balance [4-7] model of the network to compute the optimal biomass yield under conditions in which each of seven chemical compounds (called environments) – acetate, glucose, glycerol, lactate, lactose, pyruvate, and succinate – respectively served as the main metabolic substrate. The considered environments are qualitatively different, as indicated by differences in the specific growth rate μ of *E. coli* in each environment; the rank-ordering of the environments with respect to quality is as follows: $\mu_{glucose} > \mu_{glycerol} > \mu_{lactate} > \mu_{pyruvate} > \mu_{succinate} > \mu_{acetate}$ [16]. Note that the specific growth rate in lactose was not measured in [16], so we are unable to precisely determine its location in the rank-ordering of environments. Nevertheless, it is well known that *E. coli* generally prefer glucose when grown in environments that contain a mixture of both glucose and lactose (see, e.g., experimental results in [17]) – glucose is likely a better metabolic substrate than lactose.

Before we begin presenting our results, we find it useful to put the results into perspective. The structure of an organismal GPM changes on both ecological and evolutionary time scales; changes to the GPM's structure may result from, among other factors, changes to the environment and the outcomes of interactions among individuals within a population. For a given GPM, our ability to make meaningful predictions about its structure by considering only a subset of the factors that determine that structure will depend on the degree of coupling between the underlying factors. The first set of results we describe below takes into account the effects of the environment on the GPM's structure, independently of population-level processes. For a particular environment, these results give insights into (static) statistical structures of the GPM, and they should be interpreted in that light. Subsequently, we show that some of these static insights are consistent with population-level simulations of the adaptation process and with analytic predictions of the relative speed of adaptation to different environments.

***Statistical structures of the GPM in different environments***

**Conditional probability of phenotype differences (PPD):** We begin by asking: how does the phenotype change as we move in genotype space, in search of genotypes associated with adaptive phenotypes? To answer this question, we computed the PPD, that is, the probability that two genotypes that are separated by a Hamming distance $d_h$ in genotype space map onto phenotypes whose fitnesses differ by $d_e$ [18] (see Materials and Methods). We computed the PPD from the probability of differences between the phenotypes of a large number of randomly chosen, viable reference genotypes and the phenotypes of genotypes sampled at Hamming distances $d_h$ (=1,…,166) from each reference genotype. We find that the PPD has a less rich structure in acetate, the poorest of the environments, than in the other six environments (see Figures 2 and S1). For example, in glucose (see Figure 2) the PPD has its maximum at very small phenotype differences when Hamming distances from the reference genotype are small (i.e., $1 \leq d_h < 30$). At Hamming distances $d_h \geq 30$ the PPD exhibits an interesting bi-modal behavior that is largely independent of $d_h$. Therefore, $d_h$ = ~30, which is equivalent to 18% of the metabolic network's genome size, can be thought of as a critical Hamming distance that marks a transition from local to global features of the distribution of the magnitude of phenotype changes that accompany changes to the genotype. In contrast, in acetate the PPD (see Figure 2) has a lower critical Hamming distance (~20) at which genotype and phenotype differences become de-correlated, and the PPD is uni-modal above this critical Hamming distance. Note that the vast majority phenotypes sampled at large Hamming distances from an arbitrary reference genotype have zero fitness. Therefore, the distribution of fitness differences at large Hamming distances reflects the expected distribution of the fitnesses of randomly sampled, viable genotypes.

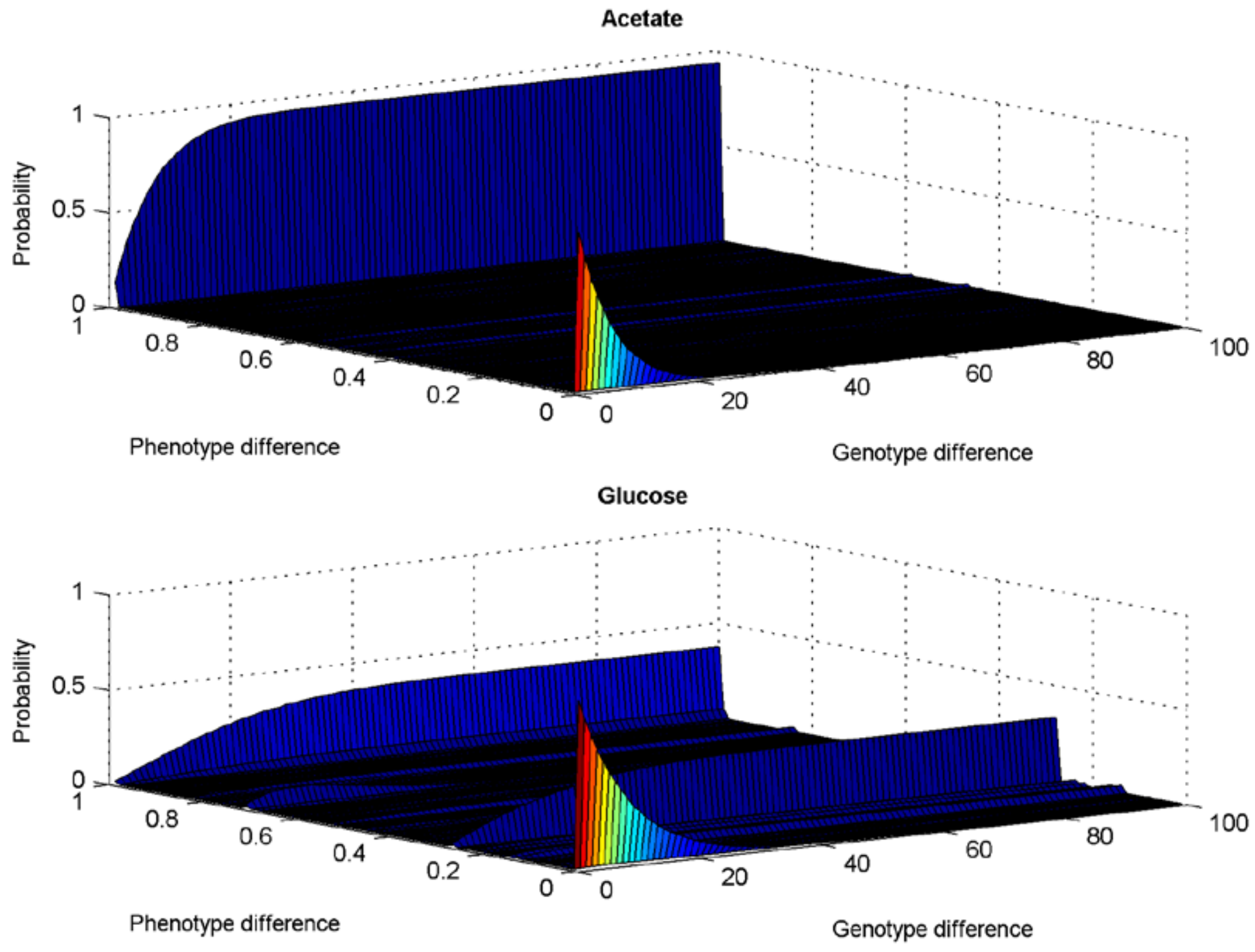

Figure 2. **Conditional probability of phenotype differences (PPD)**. The PPD was computed in acetate and glucose environments.

**Correlation length (CL) of phenotype differences:** To gain further insight into the dependence of phenotype changes on genotype changes, we computed the CL of phenotype differences, which quantifies the robustness of the phenotype to genotype changes. The longer the CL, the more robust is the phenotype. Longer CLs are also characteristic of GPMs that have a relatively smooth structure [18], in which adaptive phenotypes are more readily accessible. We found the CL to be larger in qualitatively better environments (see Figure 3). The rank-ordering of environments based on the CL ($CL_{glucose} > CL_{glycerol} > CL_{pyruvate} > CL_{lactate} > CL_{lactose} > CL_{succinate} > CL_{acetate}$) is consistent with the rank-ordering based on quality (see above), with the exception of pyruvate, which is poorer than (but is associated with a greater CL than) lactate. The CL for glycerol, lactate, and pyruvate are similar, which is consistent with the fact that both glycerol and lactose are converted into pyruvate by a small number of metabolic reactions. These results suggest that the GPM has a less rugged structure in qualitatively better environments.

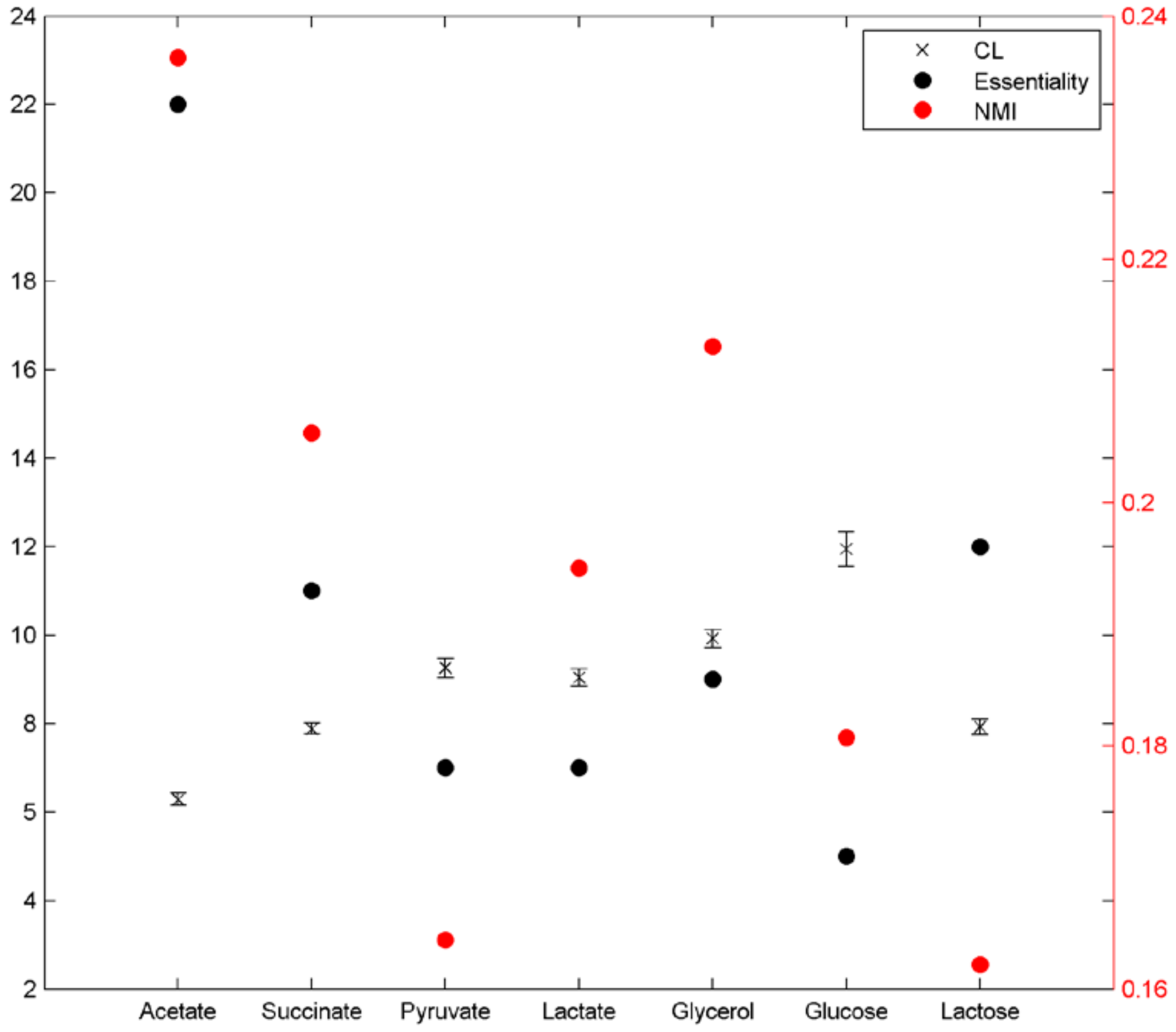

Figure 3. **Summary statistics on the structure of *E. coli's* metabolic network GPM.** Shown are the correlation length (CL) of phenotype differences, the normalized mutual information (NMI) of genotype differences relative to phenotype differences, and the number of essential

genes (essentiality) found in the metabolic network, under different environmental conditions. The environments are listed in increasing order of quality, except in the case of lactose whose position in the rank-ordering is not known precisely. The NMI was computed as described in Materials and Methods, using a mutation rate per genotype position of 0.001. Error bars indicate 95% confidence intervals.

**Normalized mutual information (NMI) of genotype differences relative to phenotype differences:** In addition to the CL, we defined another statistic called the NMI of genotype differences relative to phenotype differences (see Materials and Methods for mathematical details). The NMI quantifies the amount of information (measured in "bits") that genotype differences provide about phenotype differences, normalized by the entropy of the distribution of phenotype differences.

We will use a simple example to explain what the NMI measures. Consider a hypothetical population of individuals with known fitnesses. Suppose we wish to know the difference $d_f$ between the fitnesses of any two individuals randomly selected from the population. According to standard information-theoretic principles [19], our (average) uncertainty about the value of $d_f$ is given by the entropy of the distribution $p(d_f)$ of fitness differences between individuals found in the population: $H(d_f) = -\sum_{d_f} p(d_f)\log_2[p(d_f)]$. For example, if all individuals have the same fitness, then our uncertainty about $d_f$ will be $H(d_f)=0$ "bit" – we will know with certainty the value of $d_f$. If, on the other hand, there are $n$ possible (suitably discretized) fitness differences each of which is equally likely, then our uncertainty about $d_f$ will be maximal: $H(d_f)=\log_2(n)$ bits. Now, suppose we are told that the two individuals selected from the population have genotypes that differ by $d_g$. If there is a consistent relationship between genotype differences and phenotype differences, then knowledge of $d_g$ should decrease our uncertainty about $d_f$, that is, it should provide us with information about $d_f$. The amount of information that $d_g$ provides about $d_f$ is called the mutual information of $d_g$ relative to $d_f$ (denoted by $I(d_g; d_f)$). The NMI is the ratio of $I(d_g; d_f)$ to $H(d_f)$, that is, it measures the proportional reduction in the uncertainty about $d_f$ due to knowledge of $d_g$.

It is important to keep in mind that here we are concerned with measuring the amount of information that genotype differences convey about phenotype differences, on which natural selection acts during adaptive evolution, and not, as is often the case (e.g., see [20,21]), the amount of information that the genotype conveys about the phenotype. The NMI is lowest in environments where phenotype differences are independent of genotype differences, and it is highest (i.e., 1) in environments where phenotype differences are completely determined by genotype differences. In contrast to results based on the CL (see above), the rank-ordering of environments based on the NMI ($NMI_{acetate} > NMI_{glycerol} > NMI_{succinate} > NMI_{lactate} > NMI_{glucose} > NMI_{pyruvate} > NMI_{lactose}$) is inconsistent with the rank-ordering based on quality (see Figure 3). The results suggest that in acetate, the poorest of the environments, genotype differences could be more informative about phenotype differences than in the other environments. Note that the rank-ordering of environments based on both CL and NMI differs from the rank-ordering based on gene essentiality, a measure of the robustness of a metabolic network to gene deletions (see Figure 3). Here, gene essentiality was quantified as the number of single gene deletions that result in an unviable genotype. Also, note that part of the reason for the very low NMI in lactose is the relatively high entropy of phenotype differences computed in this environment.

**Lengths of neutral networks:** Additional information about the structure of the GPM and its potential impact on the accessibility of adaptive phenotypes is provided by the sizes of neutral networks. Neutral networks are important because they allow the search for adaptive phenotypes to proceed (by neutral drift) even if the GPM has a rugged structure. We estimated the distribution of the sizes of neutral networks by performing neutral walks on the GPM (see Materials and Methods). A neutral walk starts at a randomly chosen, viable genotype and proceeds to a random genotype located at a Hamming distance of 1 from the current genotype if: (i) the new genotype has the same phenotype as the current one and (ii) the Hamming distance between the new and starting genotypes is greater than the Hamming distance between the current and starting genotypes. A neutral walk ends when no neighbors of the current genotype satisfy these criteria. The distribution of the lengths (i.e., the Hamming distances between the final and starting genotypes) of 2000 neutral walks is shown in Figure 4. The neutral walk-lengths follow uni-modal distributions for the three environments with lowest-quality; pyruvate, acetate, and lactate. The lengths follow bi-modal distributions for all other environments. Observe that even in the environment predicted to be most rugged (acetate) neutral walks have an average length of 122.6±3.0; on average ~74% of the genotype can be altered by neutral drift without any effect on the phenotype

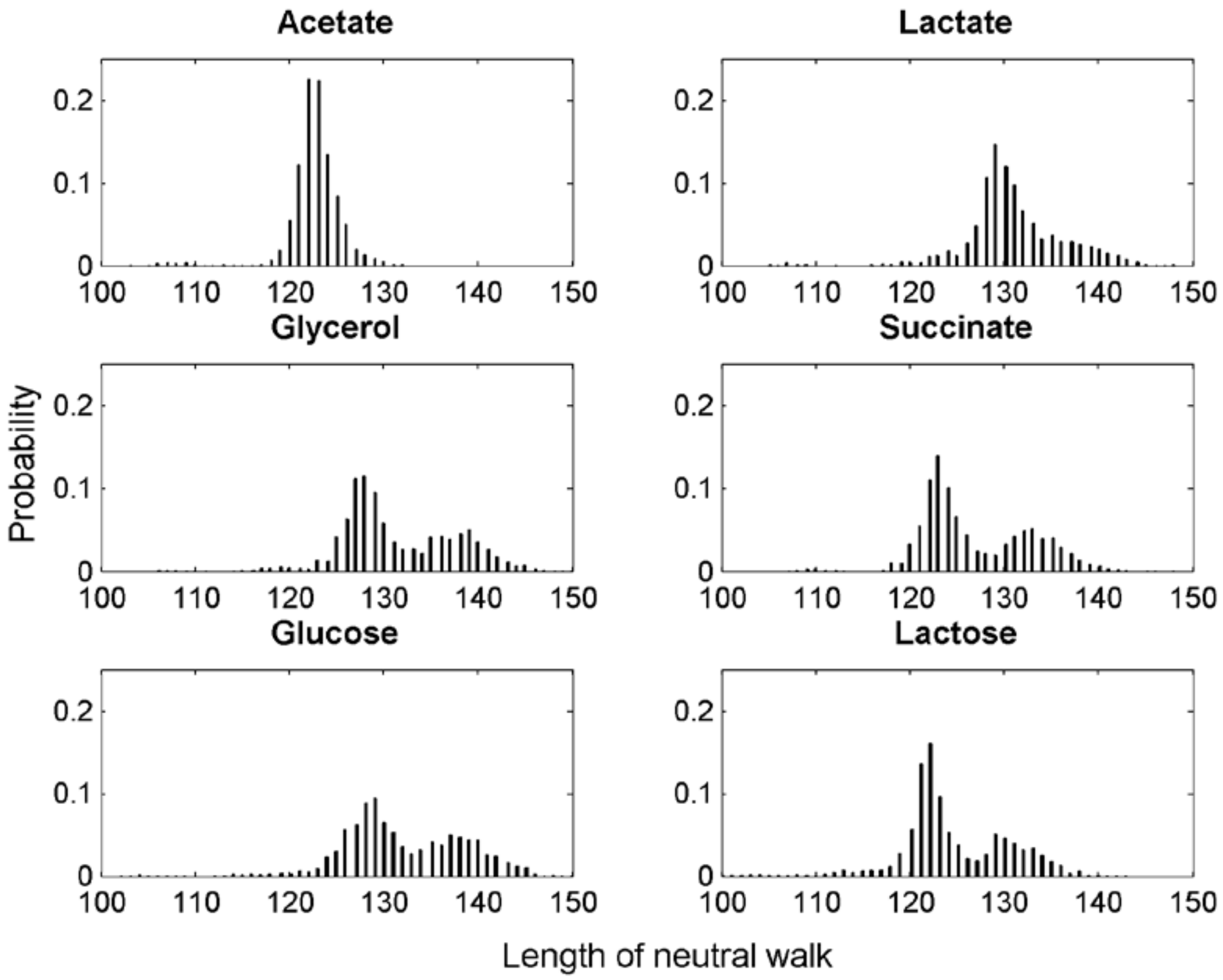

Figure 4. **Distribution of the lengths of neutral walks in different environments.** Neutral walks were performed as described in Materials and Methods. The length of a neutral walk corresponds to the Hamming distance between the final and starting genotypes associated with the walk.

### *Speed of adaption to different environments*

**Simulations-based estimates of the speed of adaptation:** In the preceding section we inferred, based on static pictures of the structure of the metabolic GPM, that the GPM has a less rugged structure in qualitatively better environments, suggesting that adaptive phenotypes could be comparatively more accessible in such environments. To gain further insight into the possible impact of environmental quality on the dynamics of adaptation, we simulated the evolutionary search for the highest-fitness phenotype in different environments. Specifically, we simulated the adaptive evolution of a population of size 1000, starting at randomly chosen genotypes with fitnesses ≤ 20% of the highest possible fitness (i.e., 1.0) (see Material and Methods for further details on these simulations). The simulations were run for a maximum of 250 generations, and they were stopped whenever the evolving population reached the target phenotype – i.e. whenever the population's mean fitness rose to within 10% of the highest possible fitness (Note that due to continual mutation, it is unlikely that an evolving population's mean fitness will equal 1.0 exactly, hence the chosen fitness cut-off). Simulations were performed in three qualitatively good environments (glucose, glycerol, and lactose), and in two comparatively poorer environments (acetate and succinate).

All evolving populations found the highest-fitness phenotype during adaptation to acetate, while 82% and 78% of the populations did so during adaptation to glycerol and succinate, respectively. In contrast, the highest-fitness phenotype was found by only 67% of populations adapting to glucose and by 63% of populations adapting to lactose. In addition, the populations that found the highest-fitness phenotype did so at a much faster rate in acetate, glycerol, and succinate than in either glucose or lactose (see Figure 5). These results are inconsistent with the expected speed of adaptation based on the CLs associated with the considered environments, but they are in agreement with the environment-specific NMIs (see Figure 3); adaptation appears to be faster in environments associated with higher NMIs.

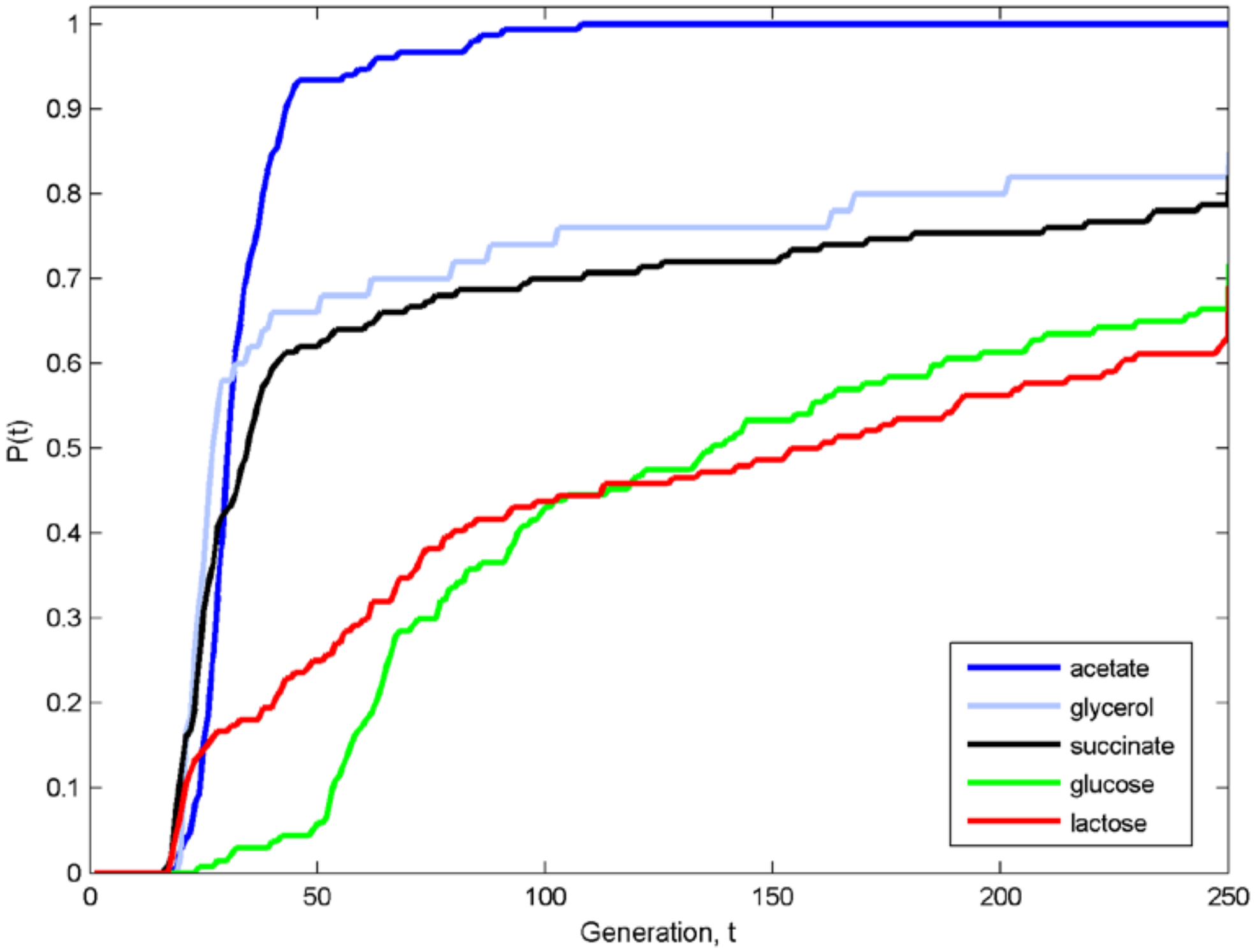

Figure 5. **Outcome of *in silico* adaptive evolution of *E. coli* populations in different environments.** The fraction $P(t)$ of evolving populations that found the highest-fitness phenotype at (or before) the $t^{\text{th}}$ generation is plotted against $t$.

**Analytic insights into the expected speed of adaptation:** The results presented above suggest the existence of a positive correlation between the NMI and the speed of adaptation. To shed additional light on this result, we now describe a simple mathematical model that makes explicit the relationship between the NMI and the speed of adaptation to a given environment, under the assumptions of Fisher's fundamental theorem of natural selection (e.g., see [22]). Consider a population consisting of $k$ "types" of individuals, with $n_i$, $i = 1, \ldots, k$, individuals belonging to the $i^{\text{th}}$ type. Let each type be characterized by its genotype, which is assumed to contain $m$ loci that have additive effects on fitness. Further, consider a hypothetical type of individuals whose fitnesses correspond to the mean fitness $\overline{w}$ of the population. Now, let the genotype of individuals of the $i^{\text{th}}$ type differ from the genotype of the abovementioned individuals, and let $a_i$ denote the corresponding fitness difference (also called the average excess in fitness; e.g., see [22]).

Mathematically, we can express the relationship between the genotype and fitness differences as: $a_i = \sum_{j=1}^{m} \alpha_j n_{ij} + \varepsilon_i$, where $n_{ij}$ denotes the number of differences occurring at the $j^{\text{th}}$ locus, and $\alpha_j$ denotes the average effect of those differences on fitness differences. We can think of $\varepsilon_i$ as the portion of fitness differences explained by the environment and other random, non-genetic factors. In general, the average effects of genotype differences will result from additive contributions of individual genes as well as interaction effects (due to, e.g., epistasis). As in the original formulation of the fundamental theorem of natural selection, we do not explicitly model the interaction effects but we include in $\varepsilon_i$ the deviation from additivity of the average effects of genotype differences on fitness differences.

The relationship between genotype and fitness differences for all types of individuals found in the population can be written as:

$$A = H * X + E, \tag{1}$$

where $A$ is a 1 by $m$ vector consisting of the $a_i$'s, $H$ a 1 by $k$ vector consisting of the $\alpha_j$'s, $X$ a $k$ by $m$ matrix whose entries are the $n_{ij}$'s, and $E$ a 1 by $m$ vector consisting of the $\varepsilon_i$'s. We let $\varepsilon_i \sim N(0, \sigma^2),\ i = 1, \ldots m$, where $N$ denotes the Gaussian distribution and $\sigma^2$ the variance. Let $p_j$ denote the probability of a difference at each position of locus $j$ and let $l_j$ denote the length of the locus. Then, for large $l_j$ we can approximate the distribution of each row of *X* by a multidimensional Gaussian with an *m* by *m* covariance matrix $\Sigma$, where the diagonal entries of $\Sigma$ are the variances of the number of differences occurring at each locus – $l_j p_j (1 - p_j)$ – and the off-diagonal entries are the covariances between the number of differences occurring at different pairs of loci. The additive genetic variance in fitness is given by $\hat{H} * \Sigma * \hat{H}^T$, where $\hat{H}$ denotes the least-squares estimate of *H*. According to standard least-squares theory, the additive genetic variance in fitness is maximized when:

$$\hat{H} = A * X^*, \tag{2}$$

where $X^*$ denotes the Moore-Penrose pseudo-inverse of *X*.

The mutual information of genotype differences relative to fitness differences is given by (e.g., see [19]):

$$I(X; A) = \frac{1}{2} \log_2 \left( 1 + \sigma^{-2} \hat{H} * \Sigma * \hat{H}^T \right), \tag{3}$$

and it is similarly maximized when $\hat{H}$ is given by (2). The mutual information increases with the additive genetic variance in fitness, which, according to Fisher's fundamental theorem of natural selection (e.g., see [22]), scales linearly with the rate of increase in fitness, under fixed environmental conditions (i.e., $\sigma^2$ fixed). Therefore, fitness should also increase with the mutual information. In other words, under given environmental conditions the speed of adaptation should be positively correlated with the mutual information of genotype differences relative to fitness/phenotype differences. More generally, normalization of the mutual information by the entropy of the distribution of fitness differences, which gives the NMI, controls for environment-specific differences in the non-genetic component of fitness, and allows comparison of the mutual information (and the expected speed of adaptation) across environments. Note that in [22], it was suggested, under certain simplifying assumptions, that the "acceleration" of the Shannon entropy is mathematically equivalent to the Fisher information, which was in turn related mathematically to the additive genetic variance in fitness. But, no explicit connection was suggested between the mutual information and the additive genetic variance in fitness, as we did here.

## Discussion

Bacterial evolution experiments have demonstrated that the environment can exert an important influence on the structure of the genotype-phenotype map (GPM). For example, Remold and Lenski [23] showed that the environment interacted synergistically with the genetic context to affect the fitness consequences of mutations introduced artificially into *E. coli* populations. Here, we asked a general question, a comprehensive investigation of which is currently only feasible by theoretical means: how does the environment affect those properties of the GPM that are important for the evolutionary accessibility of adaptive phenotypes? Four properties of the GPM were of particular interest to us: (i) the phenotypic response to genotype changes, that is, how the phenotype changes as we move in genotype space, (ii) the characteristic correlation length (CL) of phenotype differences, which measures the robustness of the phenotype to genotype changes, (iii) the normalized mutual information (NMI) of genotype changes relative phenotype changes, which quantifies the GPM's capacity to transmit information about phenotype differences, and (iv) the distribution of the lengths of neutral walks, which gives insight into an evolving population's capacity to circumvent a rugged GPM structure. We investigated the above GPM properties, using an empirical model of bacterial metabolism [4-15].

***Statistical and information-theoretic perspectives on the GPM***

We found that in all environments (except acetate) large genotype changes (> ~30) induce phenotype differences that follow an interesting bi-modal distribution. This bi-modal distribution is characteristic of the expected distribution of phenotype differences between randomly sampled genotypes, suggesting that in the considered environments the *E. coli* metabolic network maps onto two dominant clusters of similar metabolic phenotypes. In acetate, the poorest environment, the distribution of phenotype differences induced by large genotype changes was essentially uni-modal, suggesting the existence of only one dominant cluster of similar metabolic phenotypes. The CL was shorter in poorer environments, suggesting that the GPM could have a more rugged structure in such environments and, hence, it may be intrinsically more difficult to find adaptive phenotypes. Note that in poorer environments there may be fewer possibilities for re-routing fluxes through the metabolic network in order to maintain biomass yields following gene deletion; this could account for the faster decay of the correlation between biomass yields attained before and after gene deletions and, hence, the lower CLs computed in these environments.

In spite of the predicted ruggedness of the GPM in acetate, the poorest of the considered environments, very long (~74% of the genotype length) neutral walks could still be performed on the GPM, suggesting that neutral drift can alter a substantial fraction of the phenotype during evolution. In other words, a population evolving in acetate could explore large portions of genotype space by drifting on neutral networks, increasing its likelihood of discovering adaptive phenotypes. Furthermore, the NMI was largest in acetate and smallest in lactose, suggesting that the information-transmission capacity of the GPM does not necessarily increase in better environments.

***Implications for the dynamics of adaptation***

In order to gain further intuition about how qualitative changes to the environment could influence the dynamics of adaptation, we simulated the adaption of *E. coli* populations to qualitatively different environments. The speed of adaptation to a given environment was positively correlated with the NMI associated with that environment; adaptation appeared to increase with the GPM's capacity to transmit information about phenotype differences under given environmental conditions. In contrast, the relative speed of adaptation to different environments was inconsistent with expectations based on the environment-specific CLs. This suggests that the CL, and the degree of ruggedness of the GPM that it measures, may not capture enough information about features of the GPM that influence the speed of adaptation. The above results were found to be consistent with the predictions of a mathematical theory that, under the assumptions of Fisher's fundamental theorem of natural selection (e.g., see [22]), demonstrated the existence of a positive correlation between the NMI and the rate of fitness increase (i.e., the speed of adaptation). Together, the above results suggest that environmental quality could have a fundamental influence on the outcome of adaptation.

Note that previous work [24,25] showed that in a changing environment, the speed of adaptation may increase with the mutation rate and also with the propensity of point mutations to have phenotypic effects. The mathematical theory presented here provides a complementary perspective: if both the environment and the mutation rate are fixed, then the speed of adaptation may increase with the amount of information that genetic variation provides about phenotypic variation and, due to the symmetry of the mutual information, with the amount of information that phenotypic variation provides about underlying genetic variation. This suggests an intriguing connection between the "predictability" of the genetic basis of fitness increases of a particular magnitude and the rate at which such increases occur. Also, note that the NMI is, in essence, a measure of adaptation potential (or evolvability). It is applicable to a wider range of data types (both numeric and symbolic data types) than related measures of evolvability used in quantitative genetics, such as the ("narrow-sense") heritability of phenotype [26], defined as the ratio of the additive genetic variance in phenotype to the total variance in phenotype.

### *Future directions*

We conclude by pointing out some limitations of our empirical GPM model, and we discuss possible directions for future work. Firstly, our approach to analyzing *E. coli*'s metabolic network GPM did not take into account transcriptional regulation, which has been shown [27,28] to mediate dynamic microbial responses to environmental perturbations. By accounting for transcriptional regulation we would endow the metabolic network with a much richer structure that may provide additional information about the evolutionary accessibility of the network's adaptive phenotypes. Nevertheless, there is ample experimental evidence [7,10-15] that the model underlying our approach to analyzing the network is sufficient for predicting bacterial growth phenotypes in various environments. Secondly, we only considered genotype changes that turn a gene either off (i.e., deleterious changes to the gene or to its associated transcription factor) or on (as could happen when compensatory changes occur). This was motivated by practical considerations: computational prediction of graduated phenotypic consequences of genotype changes is currently not feasible on a genomic scale. Future improvements in our ability to make such predictions will allow for better modeling of metabolic network GPMs.

The GPM model we studied will add to the suite of available models (e.g., see [25,29,30]) that have enabled the investigation of important questions in evolutionary biology. In addition, the insights we presented could contribute to the understanding of evolutionary processes at both the molecular and population levels. At the molecular level, the NMI could be useful for understanding the evolvability of proteins. For example, one expects the nucleotide sequences of proteins that are particularly important for the adaptation of a pathogen to the immune response of its host (e.g., the hemagglutinin protein of influenza viruses) to occupy regions of genotype space associated with NMI values that are significantly greater than random expectations. To test this hypothesis, the NMI of the nucleotide sequence variation in a population sample of nucleotide sequences of a pathogen's protein relative to the corresponding amino acid sequence/protein structure variation can be computed. The computed NMI can be subsequently compared to the distribution of NMIs obtained from appropriately randomized (e.g., see [21,31]) versions of the original sample of nucleotide sequences to determine its statistical significance.

In addition, since the NMI affords an analytically tractable measure of evolvability, it could be useful to the mathematical investigation of the evolutionarily important relationship between evolvability and robustness (e.g., see [32]). Of particular interest is the derivation of a broadly applicable mathematical description of this relationship. Previous simulation studies of the RNA GPM (e.g., see [1,33]) showed that evolvability can increase with the robustness of RNA structures to nucleotide changes. In contrast, a recent simulation study of GPMs generated by a model gene network showed that the fraction of phenotypically consequential point mutations to a genotype of the network, which is inversely correlated with the network's robustness, increased with evolvability, during adaptation to a changing environment [25]. It is not clear whether these conflicting results can be obtained from different instantiations of the same mathematical model or whether they are fundamentally irreconcilable. Additional insight could come from mathematical investigations of simple model GPMs (e.g., see [30]) using analytically tractable measures of evolvability (e.g., the NMI) and robustness (e.g., the CL; see [18]). These investigations could yield important insights into the possible existence of general mathematical rules underlying the relationship between evolvability and robustness.

## Materials and Methods

***Central metabolic network of E. coli*:** A number of reconstructions of the *E. coli* metabolic network have been published and used to obtain important insights into ways that the bacteria organize their fluxes in order to achieve optimal growth rates in different environments (e.g., see [6-15]). For our current purposes, we sought a reconstruction that satisfied the following criteria: (i) its predictions have been validated in a rigorous manner, and (ii) it is not so complex as to preclude intensive computational analyses. Based on these criteria, we chose the central metabolic network described in [9] as a starting point. We updated the reactions (including the stoichiometries of reactants and products) based on information presented in [10]. We also updated information about the enzymes (and associated genes) that catalyze reactions found in the network based on the relevant gene-protein-reaction associations data [10]. The updated central metabolic network contains reactions catalyzed by enzymes encoded by a total of 166 genes (see Table S1). The genome of the network is defined as an ordered list of these 166 genes.

***Definition of the metabolic network's genotype and phenotype*:** A genotype of the metabolic network corresponds to a particular state of the network's genome (defined above). Mathematically, we represent the genotype as an ordered list of binary values (0 or 1), with a "1" at position *x* of the genotype indicating that the gene at position *x* of the genome is active or "on", and a "0" indicating that the gene is inactive or "off". The Hamming distance between any two genotypes is the number of differences in the on-off states of corresponding genes found in both genotypes. Each genotype defines a unique set of constraints on metabolic reaction fluxes. The phenotype of a given genotype is the maximum biomass yield that is attainable under the constraints defined by that genotype; this definition of phenotype/fitness is well grounded in experimental data (e.g., see [11-17]). The maximum biomass yield is computed by means of flux-balance analysis [4-7], under “environmental” conditions in which one of seven compounds (acetate, glucose, glycerol, lactate, lactose, pyruvate, and succinate) serves as the primary metabolic substrate. For each environment, the upper bound of the input flux through the exchange reaction for the metabolic substrate associated with that environment is set to 10, while input fluxes through all other substrates are set to zero. An upper bound of 1000 is assigned to all unconstrained input/output fluxes, except for fluxes through the exchange reactions for oxygen, carbon dioxide, and inorganic phosphate, which are constrained be less than or equal to 50. This upper bound makes oxygen non-limiting for bacterial growth in the considered environments. Note that all computed phenotypes/fitnesses were scaled so that the highest fitness computed in a given environment was equal to 1.0. In the considered environments the fitnesses of viable genotypes were $\geq \sim 1\times10^{-2}$, while the fitnesses of unviable genotypes were $\leq \sim 1\times10^{-9}$ (essentially equal to 0. No fitness values occurred between these two limits

***The genotype-phenotype map (GPM)*:** A genotype (respectively phenotype) space refers to a structural arrangement of genotypes (respectively phenotypes) based on the Hamming (respectively Euclidean) distances between those genotypes (respectively phenotypes). A GPM is a mapping from genotype space onto phenotype space. When the phenotype is fitness, as is the case in the present study, the geometric structure of the GPM is called a fitness landscape.

***Conditional probability of phenotype differences***: The probability $p(d_e|d_h)$ that two genotypes that are a separated by a Hamming distance $d_h$ in genotype space map onto phenotypes that have a phenotype difference of $d_e$ is given by [18]:

$$p(d_e \mid d_h) = \frac{n(d_e \mid d_h)}{\sum_{d_e} n(d_e \mid d_h)}, \quad (4)$$

where $n(d_e|d_h)$ denotes the number of instances when two genotypes separated by Hamming distance $d_h$ map onto phenotypes that differ by $d_e$. $p(d_e|d_h)$ is computed by the following uniform sampling algorithm [18]:

1. Choose a reference genotype at random.
2. Sample exactly $l$=10 genotypes at each Hamming distance $h$=1,2,…,165 from the reference genotype, plus the only genotype found at $h$=166.
3. Compute the phenotype/fitness (i.e., the optimal biomass yield) of each genotype sampled in step 2. Normalize the computed fitnesses by dividing by the highest-possible fitness in the current environment (this facilitates the comparison of fitnesses across environments). Calculate the absolute difference between the computed fitnesses and the fitness of the reference genotype.
4. Arrange the fitness differences computed in step 3 into $(d_e|d_h)$ bins; note that only the $d_e$ values were binned. Bins of size 0.01 were used (there were 100 bins, with right edges at 0.01, 0.02,…,1.0). Both smaller (0.001) and larger (0.05) bin sizes gave qualitatively similar distributions for $(d_e|d_h)$ (e.g., see Figure S2).
5. Repeat the above steps until convergence of $p(d_e|d_h)$.

The above algorithm converges relatively fast (i.e., $p(d_e|d_h)$ does not vary by >10% at convergence; e.g., see Figure S3). We performed 2000 repetitions of the algorithm, generating $3.3\times10^6$ data points in the process.

***Correlation function of phenotype differences***: The correlation function describes, for example, how the similarity between the phenotype of a given genotype and that of an ancestral genotype decays as the two genotypes diverge. The correlation function of phenotype differences can be obtained directly from the quantity $n(d_e \mid d_h)$ computed in the preceding section. It is given by [18]:

$$c(h)=1-\frac{\sum_{d_e}(d_e)^2 n(d_e \mid d_h)}{\sum_{d_h}\sum_{d_e}(d_e)^2 n(d_e \mid d_h)p(d_h)}, \quad (5)$$

where

$$p(d_h)=\binom{G}{d_h}\alpha^{-G} \quad (6)$$

is the probability that the Hamming distance between two genotypes sampled randomly from genotype space equals $d_h$. In (6), $G$=166 is the genotype length and $\alpha=2$ is the number of symbols in the genotype alphabet (1/α is the probability that two genotypes uniformly sampled from genotype space differ at a particular genotype position). The correlation length, CL, is obtained by fitting $c(h)$ to exp(-$d_h$/CL), via minimization of the sum of squared errors.

Note that the above statistical methods are applicable to any mapping from a combinatorial set (e.g., the set of possible metabolic genotypes, which consist of sequences defined on a binary alphabet) onto a set consisting of either continuous- (e.g., the set of possible metabolic phenotypes/maximum biomass yields) or discrete-valued entities, whenever both the domain and range of the mapping are equipped with appropriate metrics (e.g., $d_h$ and $d_e$). The applicability of the methods does not depend on the specifics (e.g., folding thermodynamics, in the case of RNA GPMs, or flux-balance analysis, in the case of our metabolic network GPM) of the mapping under consideration.

***Mutual information of genotype differences relative to phenotype differences*:** The mutual information is a standard information-theoretic quantity [19]. The mutual information of a random variable *Y* relative to another random variable *X* quantifies the difference between the entropy *H*(*X*) of the probability distribution *p*(*X*) of *X* and the expected value of the entropy *H*(*X*|*Y*) of the conditional probability distribution *p*(*X*|*Y*). In other words, it measures the difference between the total uncertainty about *X* and the uncertainty about *X* that remains after we know *Y* (i.e., the uncertainty that is eliminated by knowledge of *Y*). In the current context, the mutual information measures the amount of information that genotype differences (corresponding to *Y*) provide about phenotype differences (corresponding to *X*) and vice-versa. We compute the mutual information as follows:

$$I(Y;X) = H(X) - H(X \mid Y) \tag{7}$$

$$= \sum_{d_e} p(d_e)\log_2[p(d_e)] - \sum_{d_h} p(d_h)\sum_{d_e} p(d_e \mid d_h)\log_2[p(d_e \mid d_h)], \tag{8}$$

where

$$p(d_e \mid d_h) = \frac{n(d_e \mid d_h)}{\sum_{d_e} n(d_e \mid d_h)} \text{ and} \tag{9}$$

$$p(d_e) = \sum_{d_h} p(d_e \mid d_h)p(d_h). \tag{10}$$

Both $n(d_e \mid d_h)$ and $p(d_h)$ are defined above. We normalize the mutual information by $H(X)$ – obtaining the NMI – in order to control for differences in the entropy of $p(X)$ in different environments. Note that the mutual information is symmetric: $I(X;Y) = I(Y;X)$.

When computing the NMI, $c$=1/α, the probability that two genotypes randomly sampled from an evolving population differ at a particular genotype position (see Eqn. 6), can be estimated from either the population's actual genetic variation (its standing genetic variation) or its potential genetic variation (its genetic variability). In the latter case, $c$ will depend on the assumed model of evolution. For example, consider a population of $N$ haploid individuals evolving from a common binary, ancestral genotype. Assuming that: (i) the mutation rate per genotype position $p$ is constant, (ii) mutations at different genotype positions segregate independently of each other, and (iii) a particular genotype position changes at most once, the distribution of the number of genotype positions that have changed $i$ times in the population, $0 < i < N$, can be approximated by a Poisson distribution with mean [34]:

$$F(i) = \int_0^1 f(q; N, p, s) \binom{N}{i} q^i (1-q)^{N-i} \,\mathrm{d}q\,, \tag{11}$$

where

$$f(q; N, p, s) = p \frac{1 - e^{-2Ns(1-q)}}{1 - e^{-2Ns}} \frac{\mathrm{d}q}{q(1-q)} \tag{12}$$

is the steady-state density of changed genotype positions with frequency $q$ (equivalently, the transient distribution of the frequency of changes to a particular genotype position), and $s$ is the average selection coefficient of genotype changes (Note that Eqn. (12) differs from the equivalent equation found in [34] by a factor of 2 because here we are dealing with haploid individuals). It follows from (11) that:

$$c = 2\prod_{i=1}^{N-1} e^{-F(i)}\left(1 - \prod_{i=1}^{N-1} e^{-F(i)}\right). \qquad (13)$$

In this work, we estimated the value of $c$ using $N$=1000 and $p$=0.001, corresponding, respectively, to the population size and mutation rate used in our simulations of adaptive evolution (see below). In the absence of information about the average selection coefficient of changes to our metabolic network genotypes, we used the estimate of $s$=0.02 previously reported for beneficial mutations occurring in evolving populations of *E. coli* [35]. We used the estimated value of $c$, together with Eqns. (6,8-10), to compute the NMI under different environmental conditions. The rank-ordering of environments based on the computed NMI was qualitatively similar for small values of $p$ (0.001 and 0.0001), but it was different for values of $p$ that are close to the reciprocal of the genome size (0.005 and 1/$G$) (see Figure S4).

***Neutral walks***: A neutral walk proceeds as follows [18]:

1. A "walker" starts at an initial, randomly chosen viable genotype, $x$. The walk length, $L$, and the current genotype, $y$, are initialized to 0 and $x$, respectively.
2. A genotype, $z$, is chosen randomly from among the genotypes that are a Hamming distance of 1 away from $y$.
3. The walker moves to $z$ if (i) $z$ has the same phenotype as does $x$, and (ii) the Hamming distance between $x$ and $z$ is greater than $L$. If both (i) and (ii) are satisfied, then $y$ is set to $z$ and $L$ is incremented by 1.
4. Steps 2 and 3 are repeated until it becomes impossible for the walker to move further.

***In silico simulation of adaptive evolution*:** We ran 100 *in silico* simulations of adaptive evolution of bacterial populations in each considered environment. We initialized each simulation with 1000 genetically identical individuals. The fitnesses of all genotypes were scaled such that the fittest genotype in each environment had a fitness of 1.0. The initial genotype was chosen at random subject to the constraint that it was (i) viable, and (ii) its fitness was $\leq 0.2$. We stopped each simulation when either (i) the mean fitness of the evolving population was within 10% (i.e., $\geq 0.9$) of the highest possible fitness, or (ii) the number of simulated generations was $\geq 250$; one generation equals ~20 minutes of physical time. The evolutionary dynamics were simulated by the following algorithm, which is similar in its essential features to algorithms used in [1,2]: in each generation the genome of each individual is replicated, with probability proportional to its fitness, and with fidelity equal to $1-p$, where $p=0.001$ is the mutation rate per genotype position. After replication is completed, individuals are randomly removed from the population until the population reaches its pre-replication size. Note that the mutation rate was chosen so that it is high enough to allow adaptation to occur quickly on the time scale of our simulations, but small enough so that the expected number of mutations per replication $pG<1$, where $G=166$ is the genotype length.

Computer implementations of the methods and algorithms described above are available upon request.

**Acknowledgements**: We are grateful to Simon Levin, Ned Wingreen, and three anonymous reviewers for very constructive comments on an earlier version of this manuscript; Michael Desai for very helpful discussions; and the laboratory of Bernhard

Palsson at the University of California, San Diego, CA, for providing access to the latest reconstruction of the *E. coli* metabolic network. A portion of this work was completed while W.N. was a visitor at the Kavli Institute for Theoretical Physics, Santa Barbara, CA.

## Supplemental table and figures

**Table S1. Reactions found in the *E. coli* central metabolic network analyzed in this study**

| Reaction | Protein/ protein complex | ORFs | Description |
|---|---|---|---|
| ACCOA + OA --> COA + CIT | GltA | b0720 | Citric acid cycle |
| CIT --> OA + AC | CitDEF | b0615,b0616,b0617 | |
| MAL --> OA | Mqo | b2210 | |
| CIT <--> ICIT | AcnA | b1276 | |
| CIT <--> ICIT | AcnB | b0118 | |
| ICIT + NADP <--> CO2 + NADPH + AKG | Icd | b1136 | |
| AKG + NAD + COA --> CO2 + NADH + SUCCOA | LpdA and SucAec and SucBec | b0726,b0727,b0116 | |
| SUCCOA + ADP + PI <--> ATP + COA + SUCC | Frd | b0728,b0729 | |
| FUM <--> MAL | FumA | b1612 | |
| FUM <--> MAL | FumB | b4122 | |
| FUM <--> MAL | FumC | b1611 | |
| MAL + NAD --> NADH + OA | Mdh | b3236 | |
| GLC + ATP --> G6P + ADP | Glk | b2388 | Glycolysis/ gluconeogenesis |
| G6P <--> F6P | Pgi | b4025 | |
| F6P + ATP --> FDP + ADP | PfkA | b3916 | |
| F6P + ATP --> FDP + ADP | PfkB | b1723 | |
| FDP --> F6P + PI | Fbp | b4232 | |
| FDP <--> T3P1 + T3P2 | FbaA | b2925 | |
| FDP <--> T3P1 + T3P2 | FbaB | b2097 | |
| FDP <--> T3P1 + T3P2 | B1773 | b1773 | |
| T3P2 <--> T3P1 | Tpi | b3919 | |
| T3P1 + PI + NAD <--> NADH + 13PDG | GapA | b1779 | |
| FDP --> F6P + PI | GlpX | b3925 | |
| 13PDG + ADP <--> 3PG + ATP | Pgk | b2926 | |
| 3PG <--> 2PG | GpmB | b4395 | |
| 3PG <--> 2PG | GpmA | b0755 | |
| 3PG <--> 2PG | YibO | b3612 | |
| 2PG <--> PEP | Eno | b2779 | |
| PYR + ATP --> PEP + AMP + PI | Ppsa | b1702 | |
| PEP + ADP --> PYR + ATP | Pykf | b1676 | |
| PEP + ADP --> PYR + ATP | Pyka | b1854 | |
| PYR + COA + NAD --> NADH + CO2 + ACCOA | AceEec and AceFec and LpdA | b0114,b0115,b0116 | |
| G6P + NADP <--> D6PGL + NADPH | Zwf | b1852 | Pentose phosphate pathway |
| D6PGL --> D6PGC | PGL | b0767 | |
| D6PGC + NADP --> NADPH + CO2 + RL5P | Gnd | b2029 | |
| RL5P <--> R5P | RpiA | b2914 | |
| RL5P <--> R5P | RpiB | b4090 | |
| RL5P <--> X5P | Rpeec | b3386 | |
| RL5P <--> X5P | SgcE | b4301 | |
| R5P + X5P <--> T3P1 + S7P | TktA | b2935 | |
| X5P + E4P <--> F6P + T3P1 | TktB | b2465 | |
| T3P1 + S7P <--> E4P + F6P | TalB | b0008 | |
| R5P + X5P <--> T3P1 + S7P | TalA | b2465 | |
| X5P + E4P <--> F6P + T3P1 | TktA | b2935 | |
| T3P1 + S7P <--> E4P + F6P | TktB | b2464 | |
| OA --> CO2 + PYR | Eda | b1850 | |
| ACCOA + 2 NADH <--> ETH + 2 NAD + COA | AdhE | b1241 | Pyruvate metabolism |
| PYR + COA --> ACCOA + FOR | PflC | b3951,b3952 | |
| PYR + COA --> ACCOA + FOR | TdcE | b3114 | |
| PYR + COA --> ACCOA + FOR | PflA | b0902,b0903 | |
| ACCOA + PI <--> ACTP + COA | Pta | b2297 | |
| ACCOA + PI <--> ACTP + COA | EutD | b2458 | |
| ACTP + ADP <--> ATP + AC | AckA | b2296 | |
| ACTP + ADP <--> ATP + AC | PurT | b1849 | |
| ACTP + ADP <--> ATP + AC | TdcD | b3115 | |
| ATP + AC + COA --> AMP + PPI + ACCOA | Acs | b4069 | |
| OA + ATP --> PEP + CO2 + ADP | Pck | b3403 | Anaplerotic reactions |
| PEP + CO2 --> OA + PI | Ppc | b3956 | |
| MAL + NADP --> CO2 + NADPH + PYR | Mae | b2463 | |
| MAL + NAD --> CO2 + NADH + PYR | Sfc | b1479 | |
| ICIT --> GLX + SUCC | AceA | b4015 | |
| ACCOA + GLX --> COA + MAL | AceB | b4014 | |
| ACCOA + GLX --> COA + MAL | GlcB | b2976 | |
| PPI --> 2.00 PI | Ppa | b4226 | |
| PPPI --> PI + PPI | PpxA | b2502 | |
| PPI --> 2.00 PI | PpxB | b2502 | |
| PPPI --> PI + PPI | SureEA | b2744 | |
| PPI --> 2.00 PI | SureEB | b2744 | |
| NADH + Q --> NAD + QH2 + 3 HEXT | Nuo | b2276,b2277,b2278,b2279,b2280,b2281,b2282,b2283,b2284,b2285,b2286,b2287,b2288 | Oxidative phosphorylation |
| NADH + Q --> NAD + QH2 | Ndh | b1109 | |
| FOR + Q --> QH2 + CO2 + HEXT | Fdoec | b3892,b3893,b3894 | |
| FOR + Q --> QH2 + CO2 + HEXT | Fdn | b1474,b1475,b1476 | |
| GL3P + Q --> T3P2 + QH2 | GlpD | b3426 | |
| GL3P + Q --> T3P2 + QH2 | GlpA | b2241,b2242,b2243 | |
| QH2 + 0.5 O2 --> Q + 2.5 HEXT | CyoA | b0429,b0430,b0431,b0432 | |
| PYR + NADH <--> NAD + LAC | Dld | b2133 | |
| PYR + NADH <--> NAD + LAC | Ldh | b1380 | |
| LAC + Q --> 1 PYR + 1 QH2 | Dld | b2133 | |
| QH2 + 0.5 O2 --> Q + 2 HEXT | CbdAB | b0978,b0979 | |
| NADPH + Q --> NADP + QH2 | MdaB | b0978,b0979 | |
| PYR + Q --> AC + CO2 + QH2 | PoxB | b0871 | |
| QH2 + 0.5 O2 --> Q + 4 HEXT | CydA | b0733,b0734 | |
| NADPH + NAD --> NADP + NADH | Pnt | b1602,b1603 | |
| NADP + NADH + 2 HEXT <--> NADPH + NAD | SthA | b3962 | |
| ATP <--> ADP + PI + 4 HEXT | AtpF0, AtpF1, AtpI | b3736,b3737,b3738,b3731,b3732,b | |
| | | 3733,b3734,b3735,b3739 | |
| ATP --> ADP + PPPI | PpkA | b2501 | |
| ATP + PI --> ADP + PPI | PpkB | b2501 | |
| NADPH + NAD --> NADP + NADH | SthA | b3962 | |
| SUCC + Q --> QH2 + FUM | Sdh | b0721,b0722,b0723,b0724 | |
| O2xt <--> O2 | O2TXR | | |
| CO2xt <--> CO2 | CO2TXR | | |
| ATP --> ADP + PI | ATPM | | |
| LCTS --> GLC + bDGLAC | LacZ | b0344 | Alternate carbon metabolism and related reactions |
| LCTS --> GLC + bDGLAC | BglX | b2132 | |
| bDGLAC <--> GLAC | GALM1R | b0756 | |
| bDGLC <--> GLC | GALM2R | b0756 | |
| GLAC + ATP <--> GAL1P + ADP | GalK | b0757 | |
| GAL1P + UTP <--> PPI + UDPGAL | GalT | b0758 | |
| GL + ATP --> ADP + GL3P | GlpK | b3926 | |
| GL3P + NADP <--> T3P2 + NADPH | GpsA | b3608 | |
| RIB + ATP --> R5P + ADP | RbsK | b3752 | |
| UDPGAL <--> UDPG | GalE | b0759 | |
| UTP + G1P <--> PPI + UDPG | GalUec | b1236 | |
| G1P <--> G6P | Pgmec | b0688 | |
| G1P <--> G6P | YqaB | b2690 | |
| ATP + AMP --> 2 ADP | Adk | b0474 | Nucleotide Salvage Pathway |
| 41.25 ATP + 3.54 NAD + 18.22 NADPH + 0.20 G6P + 0.07 F6P + 0.89 R5P + 0.36 E4P + 0.12 T3P1 + 1.49 3PG + 0.51 PEP + 2.83 PYR + 3.74 ACCOA + 1.78 OA --> 3.74 COA + 1.07 AKG + 41.25 ADP + 41.250 PI + 3.54 NADH + 18.22 NADP + Biomass | | | Biomass reaction |
| Biomass + 41.25 ATP --> 41.25 ADP + 41.25 PI | | | Growth |
| Transport reactions | | | |
| FORxt <--> FOR | FocA | b0904 | Formate transport via |
| | | | diffusion |
| LCTSxt + HEXT <--> LCTS | LacY | b0343 | Lactose transport via proton symport |
| FORxt <--> FOR | FocB | b2492 | Formate transport via diffusion |
| ETHxt + HEXT <--> ETH | ETHUPR | | |
| SUCCxt + HEXT <--> SUCC | DctA | b3528 | Succinate transport via proton symport 2 H |
| SUCCxt + HEXT <--> SUCC | DcuB | b4123 | Succintate transport via proton symport 3 H |
| SUCCxt + HEXT <--> SUCC | DcuA | b4138 | Succintate transport via proton symport 3 H |
| SUCC --> SUCCxt + HEXT | DcuC | b0621 | Succintate transport via proton symport 3 H |
| PYRxt + HEXT <--> PYR | PYRUPR | | |
| PIxt + HEXT <--> PI | PitA | b3493 | Phosphate reversible transport via symport |
| PIxt + HEXT <--> PI | PitBec | b2987 | Phosphate reversible transport via symport |
| GLCxt + HEXT --> GLC | GalP | b2943 | Glucose transport in via proton symport |
| G6Pxt + HEXT --> G6P | UhpT | b3666 | Glucose 6 phosphate transport via phosphate antiport |
| GLCxt + PEP --> G6P + PYR | Crr and PtsG and PtsH and PtsI | b2417,b1101,b2415,b2416 | Glucose transport via PEPPyr PTS |
| GLCxt + PEP --> G6P + PYR | Crr and MalX and PtsH and | b2417,b1621,b2415,b2416 | Glucose transport via |
| | PtsI | | PEPPyr PTS |
| GLCxt + PEP --> G6P + PYR | ManX and ManY and ManZ and PtsH and PtsI | b1817,b1818,b1819,b2415,b2416 | Glucose transport via PEPPyr PTS |
| GLxt --> GL | GlpF | b3927 | Glycerol transport via channel |
| RIBxt + ATP --> RIB + ADP + PI | RbsA and RbsB and RbsC and RbsDec | b3749+b3751+b3750 | Ribose transport via ABC system |
| ACxt + HEXT <--> AC | ACUPR | b4067 | Acetate reversible transport via proton symport |
| LAC <--> LACxt + HEXT | LldP | b3603 | Lactate reversible transport via proton symport |
| LAC <--> LACxt + HEXT | GlcA | b2975 | Lactate reversible transport via proton symport |
| GLCxt <--> | | | Glucose exchange |
| G6Pxt <--> | | | Glucose 6 phophate exchange |
| RIBxt <--> | | | Ribose exchange |
| GLxt <--> | | | Glycerol exchange |
| SUCCxt <--> | | | Succinate exchange |
| PYRxt <--> | | | Pyruvate exchange |
| LACxt <--> | | | Lactate exchange |
| LCTSxt <--> | | | Lactose exchange |
| FORxt <--> | | | Formate exchange |
| ETHxt <--> | | | Ethanol exchange |
| ACxt <--> | | | Acetate |
| | | | exchange |
| PIxt <--> | | | Phosphate exchange |
| CO2xt <--> | | | Carbon dioxide exchange |
| O2xt <--> | | | Oxygen exchange |

Note that reactions catalyzed by more than one protein/protein complex are listed multiple times. The data are based on information published in Covert et al. (Bioinformatics 24:2044, 2008) and the most recent reconstruction (denoted by iAF1260) of the *E. coli* metabolic network (see Feist et al., Mol Sys Biol 3:121, 2007; http://gcrg.ucsd.edu/In_Silico_Organisms/E_coli/E_coli_SBML). Metabolite names: 13PDG 1,3-bis-phosphoglycerate; 2PG 2-phosphoglycerate; 3PG 3-phosphoglycerate; AC acetate; ACCOA acetyl CoA; ACTP acetyl phosphate; ACxt acetate; ADP adenosine diphosphate; AKG alpha ketoglutaric acid; AMP adenosine monophosphate; ATP adenosine triphosphate; CIT citrate; CO2 carbon dioxide; CO2xt carbon dioxide, external; COA coenzyme A; D6PGC D-6-phosphogluconate; D6PGL D-6-phosphogluconolactone; E4P D-erythrose-4-phosphate; ETH ethanol; F6P fructose-6-phosphate; FDP fructose-1,6-biphosphate; FOR formic acid; FUM fumaric; G1P glucose-1-phosphate; G6P glucose-6-phosphate; G6Pxt glucose-6-phosphate, external; GAL1P galactose-1-phosphate; GL glycerol; GL3P glycerol-3-phosphate; GLAC galactose; GLC glucose; GLX glyoxylate; HEXT external H+; ICIT iso-citric acid; LAC D-lactate; LCTS Lactose; MAL maltose; NAD nicotinamide adenine dinucleotide; NADH nicotinamide adenine dinucleotide, reduced; NADP nicotinamide adenine dinucleotide phosphate; NADPH nicotinamide adenine dinucleotide phosphate, reduced; O2 oxygen; OA oxaloacetate; PEP phosphoenolpyruvate; PI phosphate; PPI pyrophosphoric acid; PPP inorganic triphosphate; PYR pyruvate; Q ubiquinone; QH2 ubiquinol; R5P ribose-5-phosphate; RIB ribose; RL5P ribulose-5-phosphate; S7P sedo heptulose; SUCC succinate; SUCCOA succinyl CoA; T3P1 glyceraldehyde-3-phosphate; T3P2 dihydroxyacetone phosphate; UDPG uridine diphosphate glucose; UDPGAL uridine diphosphate galactose; UTP uridine triphosphate; X5P xylulose-5-phosphate; bDGLAC beta-D-galactose; bDGLC beta-D-glucose. Additional abbreviations: ex exchange; xt extracellular.

Figure S1. Conditional probability distribution of phenotype differences computed in various environments

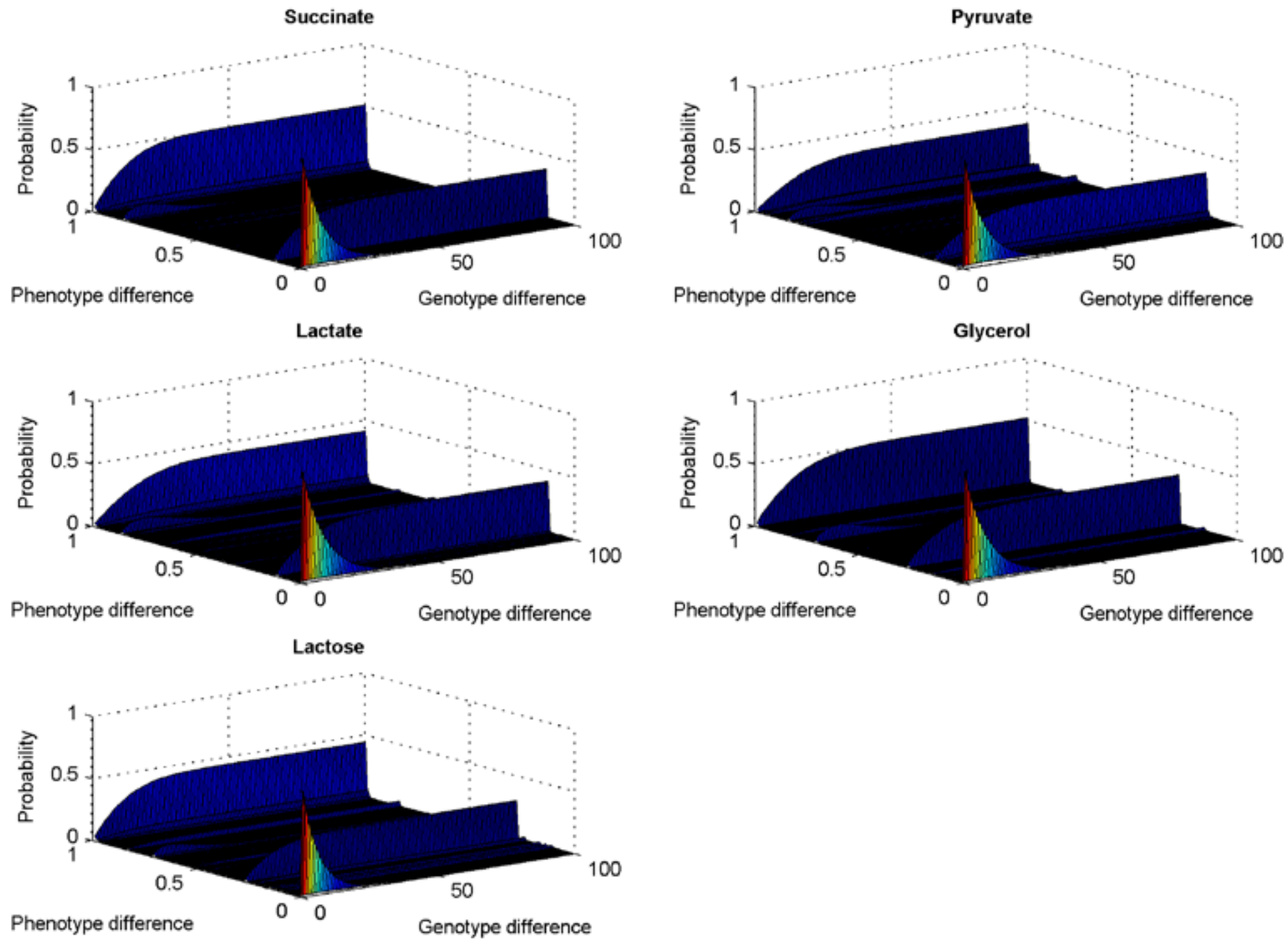

The distributions were computed as described in the main text. Phenotype differences were binned using bins of sizes 0.01.

Figure S2. Conditional probability distribution of phenotype differences

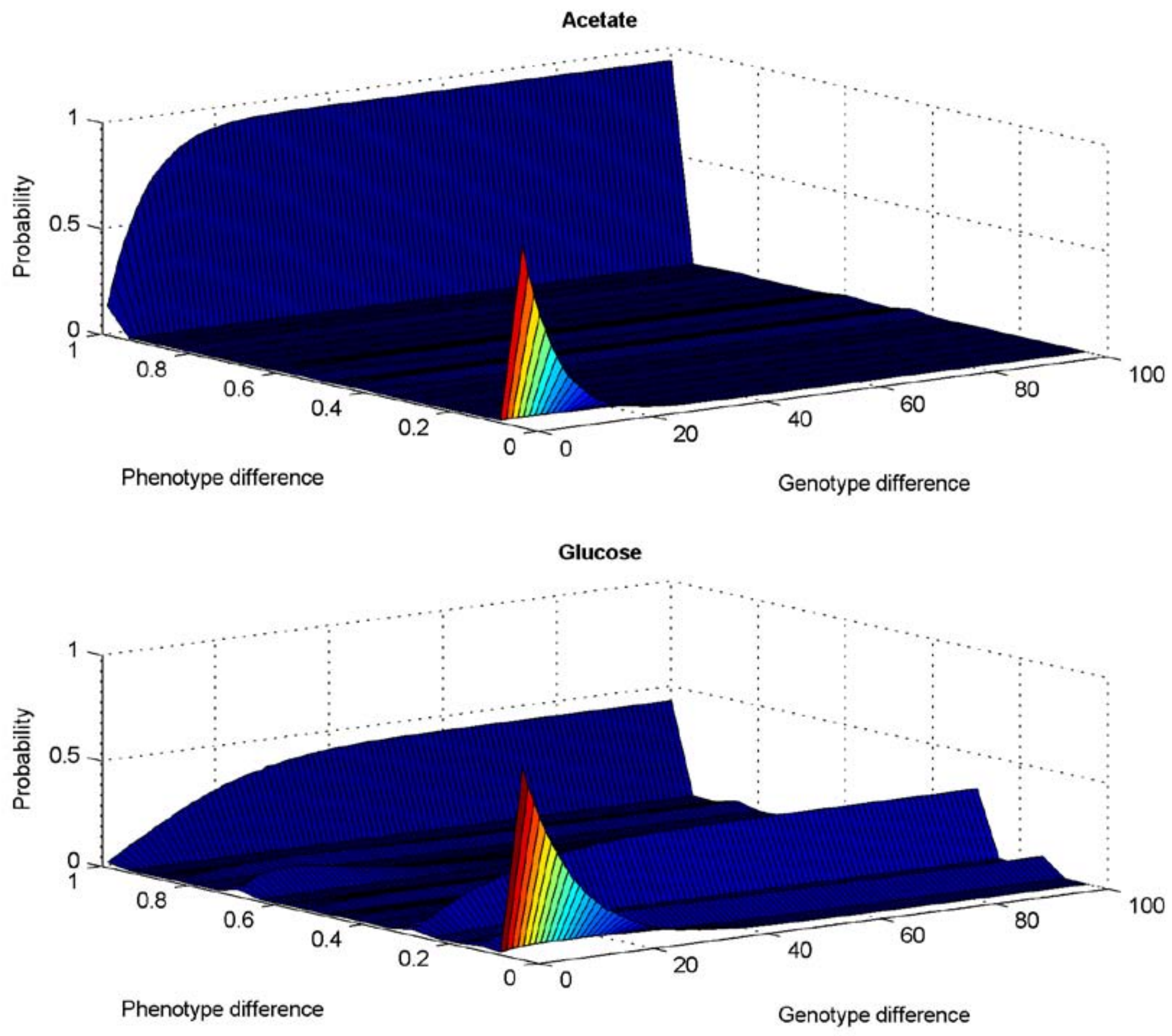

The distributions were computed as described in the main text. Phenotype differences were binned using bins of sizes 0.05.

Figure S3. Convergence of the conditional probability distribution of phenotype differences

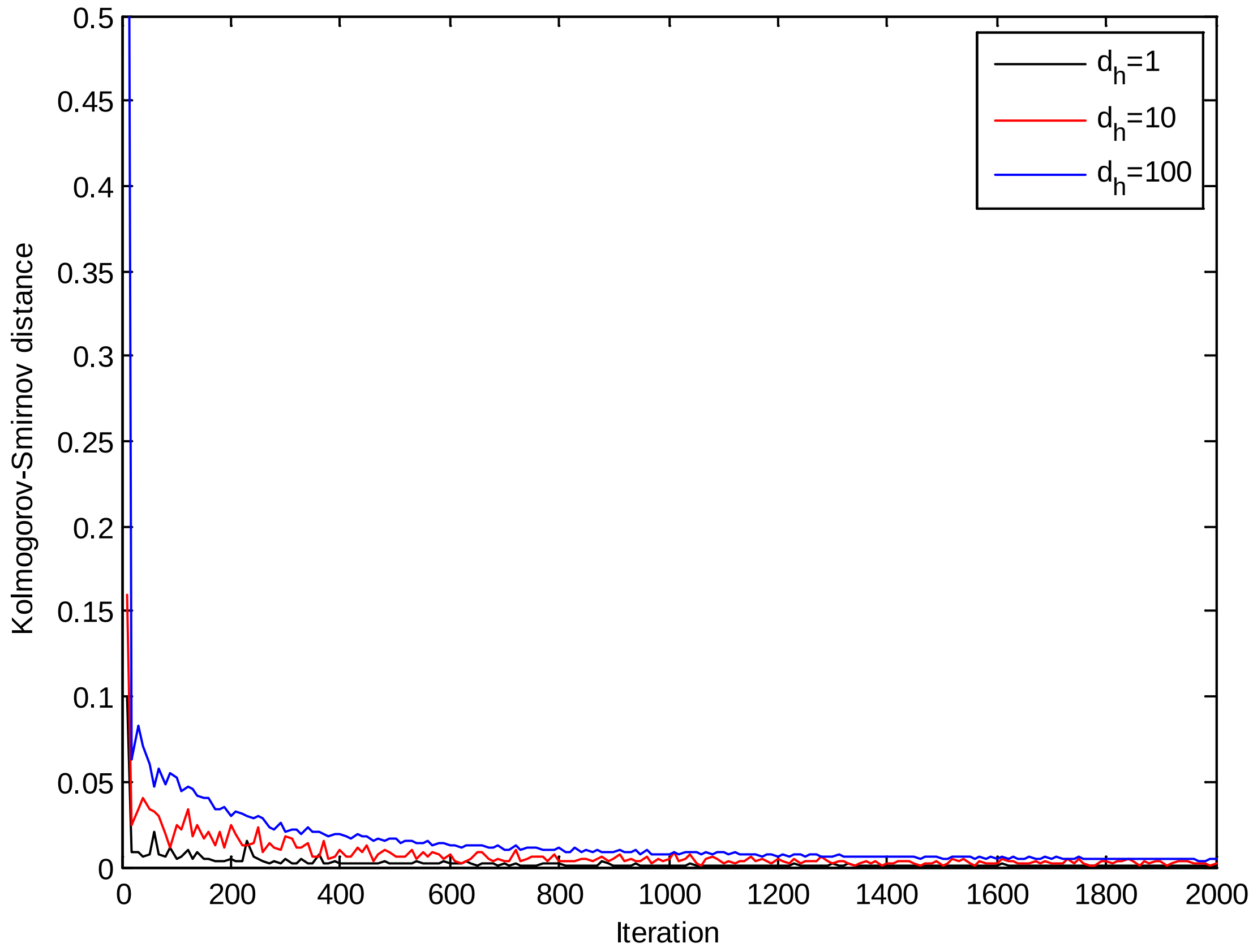

Shown is the Kolmogorov-Smirnov distance between the distribution of phenotype differences $d_e$ conditioned on genotype differences $d_h$ obtained after $t$ iterations of the uniform sampling algorithm described in the main text (denoted $p(d_e|d_h,\mathrm{t})$) and the distribution $p(d_e|d_h,t\text{-}10)$, for three values of $d_h$ spanning a wide range. The Kolmogorov-Smirnov distance is given by max{abs($p(d_e|d_h,t)$-$p(d_e|d_h,t\text{-}10)$)}. The data were collected in a glucose environment.

Figure S4. Normalized mutual information (NMI) of genotype differences relative to phenotype differences, computed in different environments

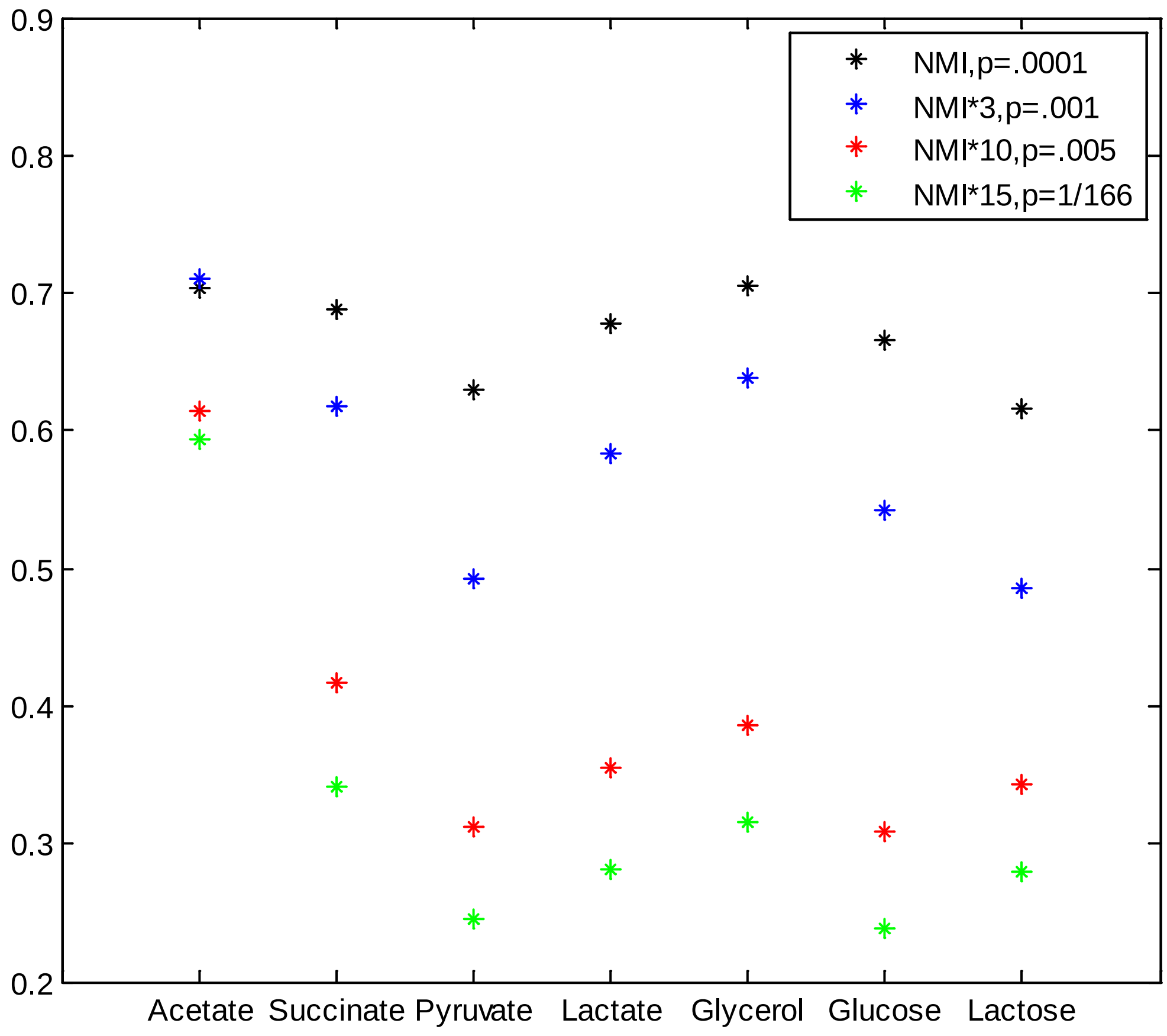

The environments are listed in increasing order of quality, except in the case of lactose whose position in the rank-ordering is not known precisely. NMI (in units of generations) was computed as described in the main text, using different values of *p*, the mutation rate per genotype position. The measurement scales of NMI values corresponding to different values of *p* were adjusted in order to facilitate their presentation on the same graph.